\documentclass[ floatfix,twocolumn]{revtex4}
\usepackage{amsmath}
\usepackage{color}
\usepackage{amssymb}
\usepackage{graphicx}
\usepackage[utf8]{inputenc}
\usepackage{textcomp}
\usepackage{color}
\usepackage{esint}

\begin{document}

\title{ Band splitting  in the altermagnet CrSb}

\author{Vladimir P.Mineev}
\affiliation{Landau Institute for Theoretical Physics, 142432 Chernogolovka, Russia}

\begin{abstract}
Altermagnets are a class of metallic magnets characterized by spin-split electron bands. Like antiferromagnets they  lack spontaneous bulk magnetisation.
The initial description of the momentum dependent spin splitting of electron bands  in altermagnets was based on the spin groups approach, which is valid when relativistic interactions are neglected. Then the  spin-orbit coupling has been taken into account according with time-reversal-odd, even-parity irreducible representations of centrosymmetric point groups formally considered by R.M.Fernandes et al, Phys.Rev.B {\bf 109}, 024404 (2024). 
The problem of electron band splitting should be discussed based on crystal symmetry of concrete altermagnet material. 
Here, the band spin-splitting in hexagonal altermagnet CrSb is established using magnetic groups formalism that  allows to find the additional spin splitting missed  in frame of exchange approximation. 

Another type of spin splitting takes place in so called hidden altermagnets. This type materials do not possess symmetry in respect of space and time inversion but are symmetric in respect of the product of these operations known as toroid symmetry. There is demonstrated the real
space spin splitting in materials with toroid symmetry.
\end{abstract}

\date{\today}

\maketitle

\section{Introduction}

The Kramers degeneracy of electron states, known in metals, is lifted in ferromagnetic materials, as well as in noncentrosymmetric metals, that is, metals with a crystal structure without mirror symmetry. This leads to a splitting of the energy of the electron bands $\delta\varepsilon({\bf k})$ and, therefore, can be detected
by measuring the difference in the de Haas-van Alphen oscillation frequencies
between the two splitting branches..
The splitting of the energy bands in noncentrosymmetric metals is determined by the spin-orbit interaction and typically ranges from several tens to several hundreds Kelvin \cite{Roth1967, Mineev2005, Terashima2008, Onuki2014, Aoki2018}.

Recently, initially in theoretical studies \cite{Noda2016, Okugawa2018, Ahn2019, Hayami2019}, a momentum-dependent spin splitting of electron bands was discovered in metallic collinear antiferromagnets in the absence of spin-orbit coupling, which also exhibit a lifted Kramers degeneracy of the electron states.
Soon after, an approach to the rigorous and systematic classification and description of nonrelativistic phases of magnetic materials was developed
\cite{Smejkal2022} based on
the spin group formalism introduced in the seminal works \cite{Brinkman1966, Andreev1980}. This approach, implemented within the so-called exchange approximation, allows one to determine possible spin structures, including symmetry operations only in spin space. A new class of magnetic materials has been introduced in which sublattices with opposite spins can be coupled by proper or improper rotations, but cannot be coupled by translation or inversion. The electronic band structure in these materials consists of momentum-dependent spin-split bands, but, like antiferromagnets, lacks spontaneous bulk magnetization.
The term "altermagnets" has been proposed for materials of this type \cite{Smejkal2022}.

Another approach to the metals with momentum dependent band splitting based on traditional symmetry classification of magnetic materials which includes both nonrelativistic and relativistic  interactions \cite{LL1984} has been developed in \cite{MineevUFN}.  Also, the role of spin-orbit interaction in relationship to the spin splitting of electron bands in centrosymmetric antiferromagnets has been considered earlier in \cite{Zunger2020}.
Dielectric antiferromagnetic materials with symmetry that includes time reversal only in combination with rotations or reflections, or none at all,
are well known as piezomagnets. 
Altermagnets according definition  introduced in \cite{MineevUFN} are metallic piezomagnets. 
Along with altermagnets that lack bulk magnetization, metallic compounds with spontaneous magnetization are also possible,
such as the ferromagnet URhGe \cite{Mineev2025}, as well as
analogs of weak ferromagnets and ferrimagnets \cite{Mineev2026}.

Recently, several research groups reported results from studies of quantum oscillations in the altermagnet CrSb \cite{Du2025, Long2026, Terashima2026}, combining magnetotransport and torque measurements with DFT+U calculations. This allowed them to identify multiple quantum oscillation frequencies originating from  spin-nondegenerate bands. The shape, position, and even symmetry of the Fermi surfaces discovered in these studies differ.

A strongly anisotropic spin-band splitting was first observed in the altermagnet MnTe, with a hexagonal crystal structure distorted by basal-plane spin ordering
\cite{Osumi2024}. This was done using photon-energy tunable ARPES in combination with first-principles calculations. This was followed by similar studies of another hexagonal altermagnet, CrSb  \cite{Reimers2024,Yang2025,Zeng2024,Ding2024,W.Li2025,C.Li2025,Liao2025}, where, however, spin ordering does not distort the hexagonal crystal structure and leads to a large g-wave spin splitting of electron bands. Spin splitting of electron bands was also registered in tunneling magnetoresistance measurements reported in \cite{X.Li2025}. Also the clear signatures of chiral spin-split magnons in CrSb have been observed in polarised neutron inelastic scattering experiments \cite{Singh2025}.

Thus, the spin-splitting of electron bands dependent from momentum direction is measurable quantity. Its theoretical description can be obtained based on symmetry considerations making use either spin-groups approach or magnetic groups one. The results of spin-groups approach for structures with different point symmetry has been presented in the Supplemental Material to the paper  \cite{Smejkal2022}. The magnetic group treatment of electron bands spin- splitting according with time-reversal-odd, even-parity irreducible representations of centrosymmetric point groups has been described in \cite{Fernandes2024}. Here we begin with short repeat of results given in this paper  in application to hexagonal ${\bf D}_{6h}$  point group. 
Then we reproduce the spin groups approach to  the band structure in CrSb. Next there will be  found the band structure corresponding   to actual magnetic symmetry of this material and stressed the  bands spin splitting difference found in frame of nonrelativistic and relativistic approach.
Final section is devoted to description of spin splitting in so called hidden altermagnets.

 \section{Electron band structure in CrSb}

\bigskip 

\begin{figure}
\includegraphics
[height=.3\textheight]
{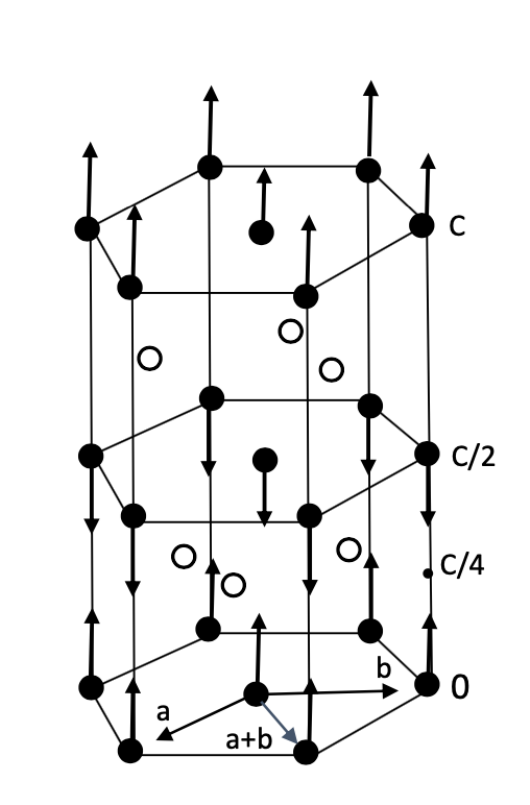}
 \caption{Crystalline and magnetic structures of CrSb. Black dots represent magnetic Cr ions, open circles represent nonmagnetic Sb ions.}
\end{figure}

The hexagonal group 
\begin{equation}
{\bf D}_{6h}\times R={\bf D}_6\times I\times R,~~~~~~~~~~~{\bf D}_6= \{ C_{n},U_n \}
\label{D}
\end{equation}
 consists of product of the group ${\bf D}_6$ and operations of space $I$ and time $R$ inversion. The group ${\bf D}_6$ consists of six rotations $C_n$ about the $\hat z$-axis by the angles $\pi n/3$ (n=0,1,...5;  $C_0=E$ is unity element) and six rotations $U_n$ by an angle $\pi$  about six axes
\begin{equation}
\hat x\cos\frac{\pi n}{6}+\hat y\sin\frac{\pi n}{6}.
\label{axes}
\end{equation}

 The electron energy dispersion for each band has the form
 \begin{equation}
\varepsilon_{\alpha\beta}({\bf k})=\varepsilon_{\bf k}\delta_{\alpha\beta}+\mbox{\boldmath$\gamma$}_{\bf k}
\mbox{\boldmath$\sigma$}_{\alpha\beta}.
\label{eqv}
\end{equation}
Here, $\mbox{\boldmath$\sigma$}=(\sigma_x.\sigma_y,\sigma_z)$ are the Pauli matrices, $ \varepsilon_{\bf k}=\varepsilon_{-{\bf k}},~\mbox{\boldmath$\gamma$}_{\bf k}=\mbox{\boldmath$\gamma$}_{-{\bf k}}$ are the even functions of momentum. 
The functions $\mbox{\boldmath$\gamma$}_{\bf k}$ describing bands spin-splitting correspond to time-reversal-odd, even-parity irreducible representations of 
the group ${\bf D}_6$. The functions of one-dimensional representations \begin{widetext}
\begin{eqnarray}
&\mbox{\boldmath$\gamma$}^{A_{1g}}_{\bf k}=a_1k_z(k_y\hat x-k_x\hat y)+a_2Im(k_x+ik_y)^6\hat z,&~~~~~~~~~~~~~~~~~~~~~~{~\bf D}_6,\\
&\mbox{\boldmath$\gamma$}^{A_{2g}}_{\bf k}=\tilde a_1k_z(k_x\hat x+k_y\hat y)+\tilde a_2Re(k_x+ik_y)^6\hat z,&~~~~~~~~~{\bf D}_6({\bf C}_6),~~{\bf C}_6=\{ C_n \},\\
&\mbox{\boldmath$\gamma$}^{B_{1g}}_{\bf k}=b_1\left[(k_x^2-k_y^2)\hat x-2k_xk_e\hat y\right ]+b_2k_zk_x(k_x^2-3k_y^2)\hat z,&~~~~~~{\bf D}_6({\bf D}_3),~~~{\bf D}_3=\{
C_{2k},U_{2k}\},\\
&\mbox{\boldmath$\gamma$}^{B_{2g}}_{\bf k}
=\tilde b_1\left [2k_xk_y\hat x+(k^2_x-k_y^2)\hat y\right ]+\tilde b_2k_zk_y(k_y^2-3k^2_x)\hat z,&~~~~{\bf D}_6({\bf D}_3^\prime),~~~{\bf D}^\prime_3=\{C_{2k},U_{2k+1}  \}.
\end{eqnarray}
\end{widetext}
are pointed in \cite{Fernandes2024}.
Here, in the right column are written  the corresponding symmetry groups of irreducible representation functions. In the parenthesis are pointed out the subgroups of the group ${\bf D}_6$ including the operations  which do not change the sign of corresponding function 
$\mbox{\boldmath$\gamma$}_{\bf k}$,  index $k=0,1,2$.  The symmetry groups of all  these states does not contain the operation of time inversion but it contains the operation of time inversion $R$ in combination with some particular proper or improper rotations.
The formal enumeration of  function of irreducible representations does not solve the problem 
of band splitting in a concrete substance with hexagonal symmetry. So, let us consider the  symmetry of CrSb.

CrSb is a metallic compound that crystallises as a hexagonal NiAs-type structure in the centrosymmetric non-symmorphic space group P63/mmc (N194).
The unit cell contains two Cr and two Sb atoms. The ordered moment in the ordered state below $T_N ~\sim 700 K$ is $\sim 3\mu_B$  parallel or antiparallel to the $\hat z$ axis, see Fig.1. The symmetry group of paramagnetic state consists of the same elements as (\ref{D}) 
but rotations with odd numbers $n=2k+1$,  $k=0,1,2$ are accompanied by a shift $t$ on half period  of crystal cell along the hexagonal axis
\begin{equation}
G_p={\bf D}^p_6\times I\times R, ~~~~~~~~~{\bf D}^p_6= \{ C_{2k},U_{2k},tC_{2k+1}, tU_{2k+1} \}
\label{G_p}
\end{equation}

 In  the ordered state (see Fig 1.) 
  some proper and improper rotations  containing the shift $t$ are accompanied by time inversion $R$ (as it should be in altermagnet state) and magnetic group of symmetry is
\begin{equation}
G_a={\bf D}^a_6\times I, ~~~~~~~~~{\bf D}^a_6= \{ C_{2k},RU_{2k},RtC_{2k+1}, tU_{2k+1} \}.
\label{alt}
\end{equation}

According to spin-groups symmetry approach \cite{Smejkal2022} 
spin part of  energy dispersion $\mbox{\boldmath$\gamma$}_{\bf k}$
is the product of scalar function of momentum  corresponding to one of irreducible representations of symmetry group of paramagnetic state (\ref{G_p})
  and axial vector $\hat z$ along hexagonal axis which is invariant in respect to symmetry operations acting only on magnetic moments that is on arrows in  Fig.1. They include:  space inversion $I$,   all rotation around $\hat z$ axis $C_z\hat z=\hat z$ and all $U_2$ rotations accompanied by operation of time reversal $UR\hat z=\hat z$,
as it should be in altermagnet state. For $B_{1g}$  representations the corresponding vector function is
\begin{equation}
\mbox{\boldmath$\gamma$}_{\bf k}^{B}=b k_z k_x(k_x^2-3k_y^2)\hat z.
\label{Bex}
\end{equation}
So, we see that according to spin-symmetry group approach the band spin-splitting possesses only momentum dependent $z$-component spin-splitting.

In frame of general  magnetic group approach the electron energy band dispersion has more general form.
Namely, 
 taking into account that time inversion changes the sign 
 ${\bf k}$-vector 
  components and axial vectors $\hat x,\hat y,\hat z$ it is easy to check that the vector function 
  \begin{equation}
\mbox{\boldmath$\gamma$}^A_{\bf k}=c_1\left[2k_xk_y\hat x+(k_x^2-k_y^2)\hat y\right ]+c_2k_zk_x(k_x^2-3k_y^2)\hat z.
\label{Aso}
\end{equation}
 corresponds to unit representation of symmetry  group of the ordered state (\ref{alt}) that is invariant in respect to all its operations.
This function determines the band spin-splitting in CrSb.
The functions $\mbox{\boldmath$\gamma$}_{\bf k}$ for other irreducible representations do not possess the symmetry corresponding to the structure of CrSb in the ordered state. 

The last term in  expression (\ref{Aso}) coincides with Eq.(\ref{Bex}) obtained by means of spin group approach \cite{Smejkal2022}. The full expression
(\ref{Aso}) does not correspond to any of band dispersion pointed in \cite{Fernandes2024} and enumerated in Eqs.(4)-(7).
 
  \begin{figure}
\includegraphics
[height=.3\textheight]
{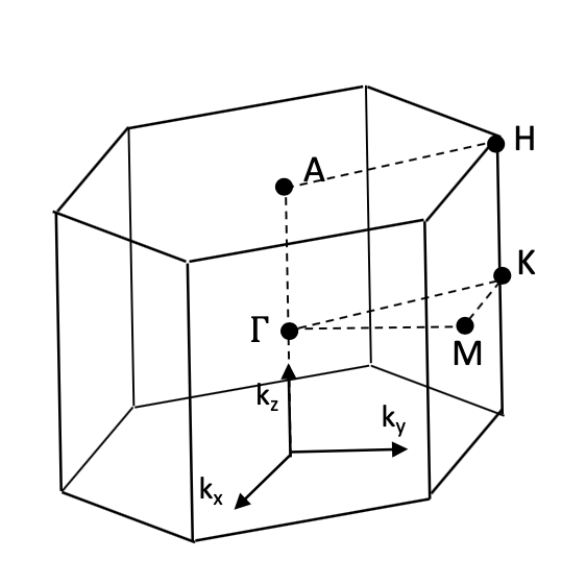}
 \caption{Brillouin zone. }
\end{figure}

 The Brillouin zone of CrSb is shown on  Fig2. The diagonalization of the matrix (\ref{eqv}) results in band dispersion laws
\begin{equation}
\varepsilon_{\pm}({\bf k})=\varepsilon_{\bf k}\pm|\mbox{\boldmath$\gamma$}_{\bf k}|,
\label{band}
\end{equation}
and the equations 
\begin{equation}
\varepsilon_{\pm}({\bf k})=\mu,
\label{band}
\end{equation}
determine the Fermi surfaces of each band split due lifted degeneracy of spin states in the altermagnet state.

 According to numerical calculations \cite{Terashima2026} the Fermi level in CrSb  is crossed by four bands split by momentum dependent internal field. There is  a band looking like the tubular sheet along  the $\Gamma A$ line and also  closed pocket around A point.  The other Fermi surface sheets are two sets of six closed pockets  located symmetrically in respect the $\Gamma A$ line at some distance from it.
The vector function $\mbox{\boldmath$\gamma$}_{\bf k}$ written above shows the momentum dependent spin direction  in each band. It should be noted that
the electron bands band structure in antiferromagnet CrSb was established in frame of non-relativistic spin-symmetry groups approach, however, the magnitude of  amplitude $b$ in Eq.(\ref{Bex}) 
determining bands splitting has more complex origin. It was found by means  ab-initio calculations including  spin-orbit coupling and has much less value than the 
typical atomic scale energy \cite{Terashima2026}.

The spin direction distribution found in \cite{Terashima2026} corresponds to expression (\ref{Bex}) obtained in exchange approximation. 
 It has  3-fold   symmetric band-splitting 
shown for instance on Fig.2 in  \cite{Terashima2026}. The spin splitting vanishes in planes passing through the line $z$ and each of  lines 
along directions (\ref{axes}) as well on plane $k_z = 0$ in reciprocal space. 

Out the exchange approximation the spin splitting on these planes is  recreated by spin-orbit coupling according to Eq.(\ref{Aso}).
The modulations of electron spin
directions in the basal plane correspond to Eq.(\ref{Aso}) Namely, to each  direction of momentum 
 $\hat{\bf k}=\hat x\cos\varphi+\hat y\sin\varphi$ corresponds the spin direction $\hat{\bf s}=\hat x\sin 2\varphi+\hat y\cos 2\varphi$. 
  The spin degeneracy is still present on the line $k_x=k_y=0$ that is $z$-axis.

In presence of external magnetic field ${\bf H}$
one must add to  $\hat \varepsilon({\bf k})$ the Zeeman term
\begin{equation}
\varepsilon_{\alpha\beta}({\bf k})~\to ~\varepsilon_{\alpha\beta}({\bf k})-\mu_{\bf k}{\bf H}\mbox{\boldmath$\sigma$}_{\alpha\beta}.
\label{10}
\end{equation}
Here, $\mu_{\bf k}$ is effective electron magnetic moment. Due to spin-orbit interaction $\mu_{\bf k}$ is  six-fold symmetric function of momentum.
Application of magnetic field decreases the symmetry of system. For instance, in case the field along  $\hat z$  axis the symmetry does not include the rotations 
$U_{2k}$. However, for any field direction the symmetry still includes space inversion $I$. 

The symmetry of  altermagnet CrSb allows piezomagnetic effect. The thermodynamic potential invariant in respect all operation group $G_a$ is
\begin{equation}
\Phi=-\lambda\left [
2\sigma_{xy}H_x
+(\sigma_{xx}-\sigma_{yy})H_y\right ]
\end{equation}
and corresponding  magnetisation arising under application of  stress $\sigma_{xx}-\sigma_{yy}$ is
\begin{equation}
M_y=-\frac{\partial \Phi}{\partial H_y}=\lambda (\sigma_{xx}-\sigma_{yy})
\end{equation}
and deformation arising under external field is 
\begin{equation}
u_{xx}=-\frac{\partial \Phi}{\partial \sigma_{xx}}=\lambda H_y.
\end{equation}

\section{Hidden altermagnetism}

The spin bands splitting   is realised in ferromagnets, noncentrosymmetric metals and altermagnets with lifted Kramers degeneracy of electron states, such that to each momentum corresponds two different electron states with different energy and different spin direction. At each point of space there are two
opposite directed density of spin (magnetisation)  non-equal each other in ferromagnets but completely compensated in noncentrosymmetric metals and altermagnets. Another interesting situation takes place in so called hidden altermagnets \cite{Guo2026,Yang2026}. 
 This type materials do not possess symmetry in respect of space inversion $I$ and time inversion $R=-i\hat \sigma^yK$ but are symmetric in respect of the product $IR$ of these operations known as toroid symmetry. Here, $K$ is the operation of the complex conjugacy.
Toroid materials have been considered theoretically in \cite{Mineev2024} and then in the review article \cite{MineevUFN}.
 In  case of toroid symmetry the Kramers theorem is valid ( see the proof in \cite{MineevUFN}) that is to each band energy $\varepsilon_{\bf k}$ corresponds two mutually orthogonal spinor Bloch functions
$\psi_{{\bf k}\alpha}({\bf r})$ and $\phi_{{\bf k}\alpha}({\bf r})=-i\sigma_{\alpha\beta}^yKI\psi_{{\bf k}\beta}({\bf r})$. One can easy  check that to each of this states corresponds the spin density
\begin{eqnarray}
{\bf S}_\uparrow({\bf r})=\psi^\dagger_{{\bf k}\alpha}({\bf r})\mbox{\boldmath$\sigma$}_{\alpha\beta}\psi_{{\bf k}\beta}({\bf r}),~~~~~~~~~~~~~~~~~~~~~~\\
{\bf S}_\downarrow({\bf r})=\phi^\dagger_{{\bf k}\alpha}({\bf r})\mbox{\boldmath$\sigma$}_{\alpha\beta}\phi_{{\bf k}\beta}({\bf r})=-\psi_{{\bf k}\alpha}(-{\bf r})\mbox{\boldmath$\sigma$}_{\alpha\beta}\psi^\dagger_{{\bf k}\beta}(-{\bf r}).
\end{eqnarray}
This means that to each spin density ${\bf S}_\uparrow$ in the point ${\bf r}$  corresponds the opposite directed density of spin ${\bf S}_{\downarrow}$
in the point $-{\bf r}$ such that the total magnetisation
\begin{equation}
\int d^3{\bf r}\left ({\bf S}_\uparrow({\bf r})+{\bf S}_\downarrow({\bf r})\right)=0
\end{equation}
is absent. One can say: here we deal with hidden magnetic properties. Another  example of toroid material is  MnAu$_2$ \cite{Fedchenko2022}. 

Unlike to altermagnets possessing by piezoelectric properties compounds with toroid symmetry possess magnetoelectric properties \cite{MineevUFN}.

\bigskip

\section{Conclusion}
The developed approach allows  to establish the  band splitting in the altermagnetic metal CrSb. 
Along with the standard description of  momentum dependent electron band spin splitting  in altermagnets 
 based on spin-group formalism valid in exchange approximation there was found 
an   additional contribution to electron band splitting  corresponding to actual crystal symmetry of  antiferromagnet CrSb.
Additional momentum dependent spin-splitting in basal plane hardly distinguishable in quantum oscillation measurements but certainly can be revealed  in neutron scattering experiments and by ARPES technic.

The structure of additional terms in band structure missed in exchange approximation corresponds to structure of piezomagnetic  thermodynamic potential determining piezomagnetic properties of CrSb
altermagnet.

Similar considerations allow establish the electron band structure in  fluorides of transient elements having  tetragonal structure. See  the preprint \cite{Mineev2026}.

At the end there was considered  the phenomenon of hidden altermagnetism  taking place in  compounds with toroid symmetry.

\bigskip

Eqs. (\ref{alt}), (\ref{Aso}) and the subsequent equations for piezomagnetic effect as they were published in PRB ${\bf 114}$, 065130 (2026)  are corrected in the present arXiv version.

\end{document}